\documentclass[reprint, aps,prd, preprintnumbers,groupedaddress,nofootinbib]{revtex4-1}
\usepackage{graphicx}
\graphicspath{{./plots/}}
\usepackage{latexsym}
\usepackage{amsfonts}
\usepackage{amssymb}
\usepackage{amsmath}
\usepackage{slashed}
\usepackage{feynmp}
\usepackage{hyperref}
\usepackage{url}
\usepackage{color}
\usepackage{cancel}
\usepackage[normalem]{ulem}
\usepackage{multirow,array}
\usepackage{dcolumn}
\usepackage{bm}

\definecolor{orangeRMA}{RGB}{255,127,0}

\begin{document}

\title{Directly Probing the Higgs-top Coupling at High Scales}

\author{Roshan Mammen Abraham}
\author{Dorival Gon\c{c}alves}
\affiliation{Department of Physics, Oklahoma State University, Stillwater, OK, 74078, USA}
\author{Tao Han} 
\author{Sze Ching Iris Leung} 
\author{Han Qin} 

\affiliation{PITT PACC, Department of Physics and Astronomy, University of Pittsburgh, 3941 O'Hara St., Pittsburgh, PA 15260, USA}

\preprint{PITT-PACC-2109}

\begin{abstract}

We explore the sensitivity to new  physics for the coupling of the Higgs boson ($h$) and top quark ($t$) at high energy scales with the process $pp\to t\bar{t}h$ at the high-luminosity LHC. This process probes the coupling in both the space-like and time-like domains at a high scale, complementary to the off-shell Higgs processes in the time-like domain. The effects from  physics beyond the Standard Model are parametrized in terms of the effective field theory framework and a non-local Higgs-top form factor. Focusing on the boosted Higgs regime in association with jet substructure techniques, we show that the present search can directly probe the Higgs-top coupling to good  precision, providing a strong sensitivity to the new physics scale. 

\end{abstract}
\maketitle

\section{Introduction}
\label{sec:intro}

The top-quark Yukawa coupling $(y_t)$ is the strongest interaction of the Higgs boson in the Standard Model (SM) with $y_t \sim 1$. Owing to its magnitude, it plays a central role in Higgs phenomenology in the SM and could be most sensitive to physics beyond the Standard Model (BSM) associated with the electroweak symmetry breaking \cite{Hill:2002ap}. It is crucial for the stability of the SM vacuum during the electroweak phase transition in the early universe~\cite{Buttazzo:2013uya,Bezrukov:2014ina}. It yields the largest quantum correction to the Higgs boson mass and can trigger the electroweak symmetry breaking in many well-motivated new physics scenarios~\cite{Carena:1993bs,Panico:2011pw,Matsedonskyi:2012ym,Pomarol:2012qf,Bellazzini:2014yua,Panico:2015jxa}. Thus, the precise  measurement of $y_t$ can be fundamental to pin down possible new physics effects.

The top-quark Yukawa coupling has been determined indirectly at the LHC from the Higgs discovery channel $gg\to h$ via the top-quark loop \cite{Aad:2019mbh}. It can also be directly measured via top pair production in association with a Higgs boson, $t\bar{t}h$. The observation of this channel was reported in 2018  by both ATLAS and CMS collaborations, with respective significances of 6.3 and 5.2 standard deviations~\cite{Aaboud:2018urx,Sirunyan:2018hoz}. These measurements confirm the SM expectation that the Higgs boson interacts with the top-quark with an order one Yukawa coupling. The high-luminosity LHC (HL-LHC) projections indicate  that the top Yukawa will be probed to a remarkable precision at the end of the LHC run, reaching an accuracy of $\delta y_t\lesssim\mathcal{O}(4)\%$~\cite{Cepeda:2019klc}.  

The current measurements are performed near the electroweak scale $Q\sim v$. If the new physics scale $\Lambda$ is significantly larger than the energy probed at the LHC, the BSM effects generally scale as $(Q/\Lambda)^n$ with $n\ge 0$ \cite{Appelquist:1974tg,Buchmuller:1985jz,Grzadkowski:2010es}, before reaching a new resonance. Therefore, it is desirable to enhance the new physics effects by exploring the high energy regime associated with the Higgs physics.
Proposals have been made recently to study the off-shell Higgs signals $gg \to h^*\to  VV$~\cite{Azatov:2014jga,Englert:2014aca,Buschmann:2014sia,Corbett:2015ksa,Goncalves:2017iub,Goncalves:2018pkt,Goncalves:2020vyn}. This process could be sensitive to  potential new physics of the $tth^*$ and $VVh^*$ interactions or a $h^*$ propagation at high energy scales $Q > v$.

In the present study, we \emph{directly} explore the Higgs-top coupling at high energy scales using the $t\bar{t}h$ production channel. For an on-shell Higgs production with high transverse momentum, this process effectively probes the top-quark Yukawa interaction at a high scale in both the space-like and time-like regimes. In contrast, the off-shell Higgs physics probes the complementary physics only in the time-like domain~\cite{Goncalves:2017iub,Goncalves:2018pkt,Goncalves:2020vyn}. As a concrete formulation, we study the BSM effects to the Higgs-top Yukawa in the Effective Field Theory (EFT) framework, focusing on two relevant higher dimensional contributions. Then, we move on to a BSM hypothesis that features a non-local momentum-dependent form factor of the  Higgs-top interaction~\cite{Goncalves:2018pkt,Goncalves:2020vyn}. This form 
factor generally captures the top Yukawa composite substructure. To combine the large event yield with a high energy physics probe, we focus on the channel with the largest Higgs decay branching fraction, $\mathcal{BR}(h\to b\bar{b})\sim 58\%$, in association with jet substructure techniques at the boosted Higgs regime. 

The rest of the presentation is organized as follows. In Section~\ref{sec:BSM}, we present the theoretical parameterization associated with the potential new physics for the Higgs-top couplings in the EFT framework and an interaction form factor. We then derive the new physics sensitivity to those interactions in Section~\ref{sec:LHC}, featuring the effects that benefit with the energy enhancement at the boosted Higgs regime. Finally, we present a summary in Section~\ref{sec:Sum}.

\section{New Physics parametrization}
\label{sec:BSM}

In this section, we describe two qualitatively different new physics parametrizations for beyond-the-Standard Model effects to the Higgs-top coupling at high energy scales. The first one considered is in the effective field theory framework by adding in a few relvant dimension-6 operators that are results from integrating out some heavy  degrees of freedom mediating the Higgs and top interactions. The second formulation is a non-local Higgs-top form factor, motivated from a strongly interacting composite theory for the Higgs and top quarks. These two forms of new physics parameterizations are quite representative in capturing the general features of the BSM couplings for the Higgs and the top quark.

\subsection{Effective Field Theory}
\label{sec:eft}

The Standard Model Effective Field Theory (SMEFT) provides a consistent bottom-up framework to search for new physics~\cite{Buchmuller:1985jz,Grzadkowski:2010es,Ellis:2020unq,Ethier:2021bye,Brivio:2019ius,Corbett:2015ksa,Biekotter:2018jzu}. In this scenario, the beyond the SM particles are too heavy to be produced on-shell. The new states can be integrated out and parametrized in terms of higher dimension operators as contact interactions  \cite{Appelquist:1974tg}. In general, the EFT Lagrangian can be written as
\begin{align}
\mathcal{L}_{\text{EFT}} = \mathcal{L}_{\text{SM}} + \sum_{i} \frac{c_i}{\Lambda^2} \mathcal{O}_i + \mathcal{O}\left(\frac{1}{\Lambda^4}\right)\,,
\end{align}
where $\Lambda$ is the scale of new physics, $\mathcal{O}_i$ are effective operators of dimension-six compatible with the SM symmetries, and $c_i$ are corresponding  Wilson coefficients. 
Higher dimensional operators can modify the existing SM interactions, as well as generate new Lorentz structures, both of which can give rise to phenomenologically relevant  energy enhancements in the scattering amplitudes.

\begin{figure}[t!]
\makebox[\linewidth][c]{%
    \centering
    \includegraphics[width=0.23\textwidth]{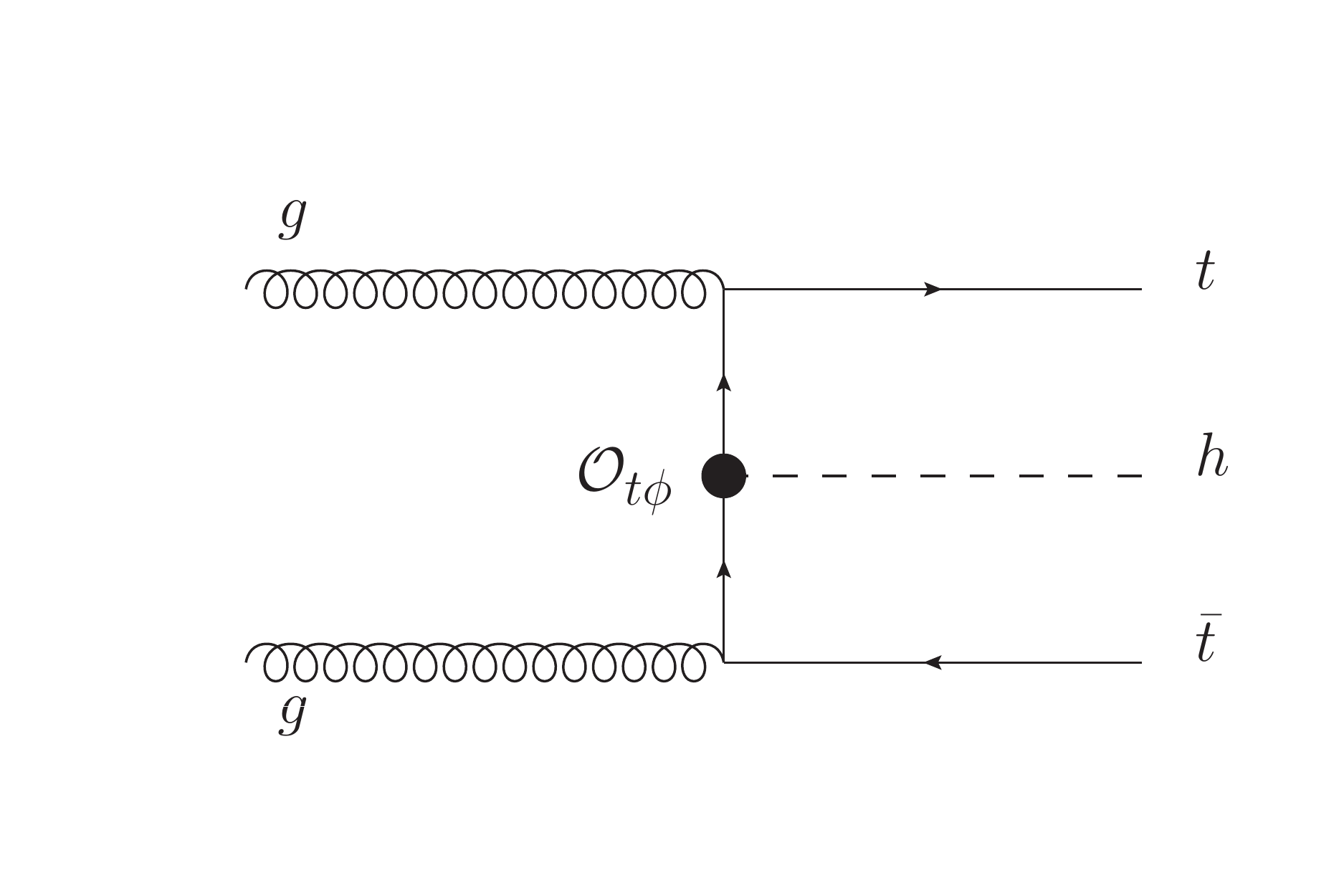} \hspace{-1.3cm}
	\includegraphics[width=0.23\textwidth]{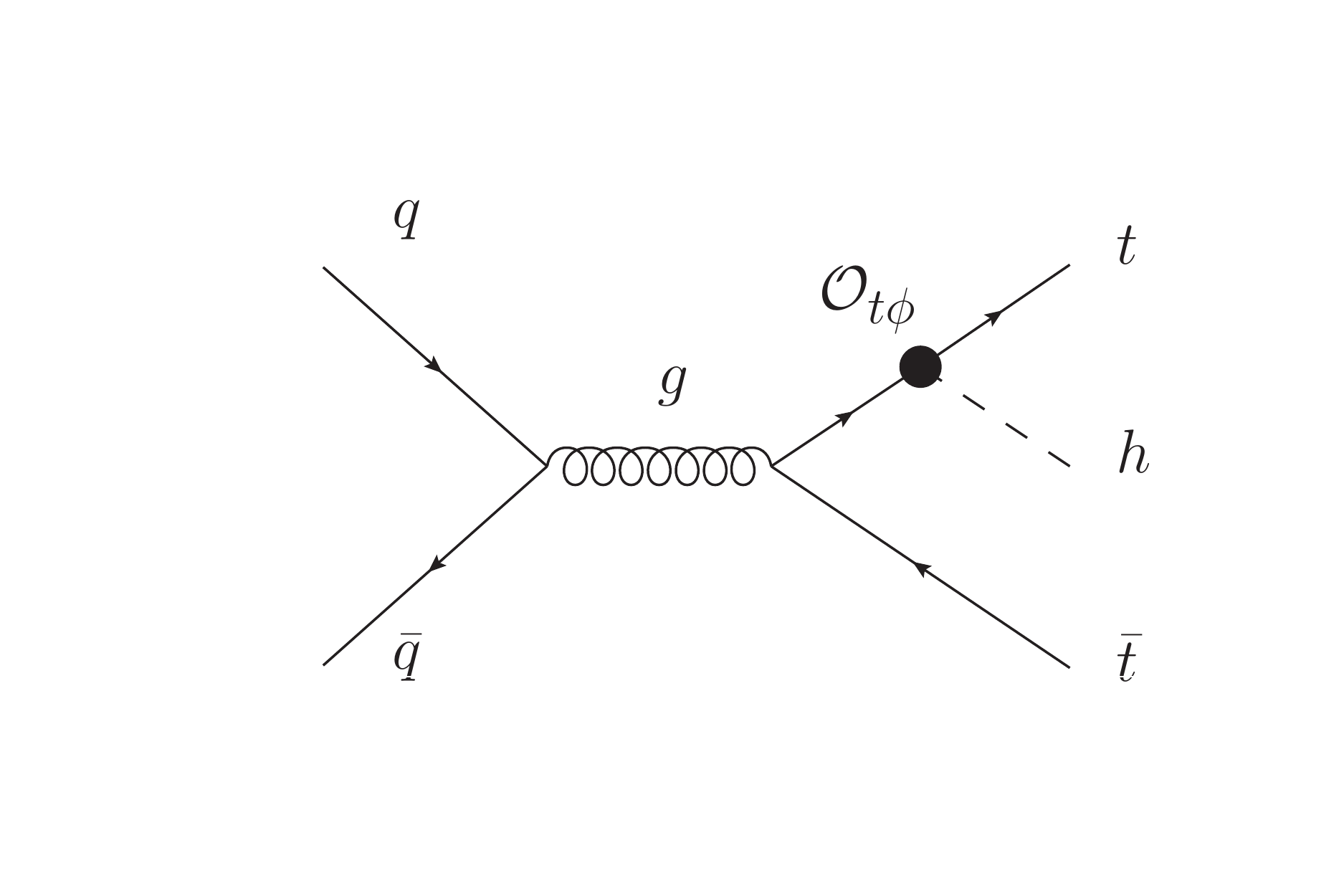} \hspace{-1.3cm}
    \includegraphics[width=0.23\textwidth]{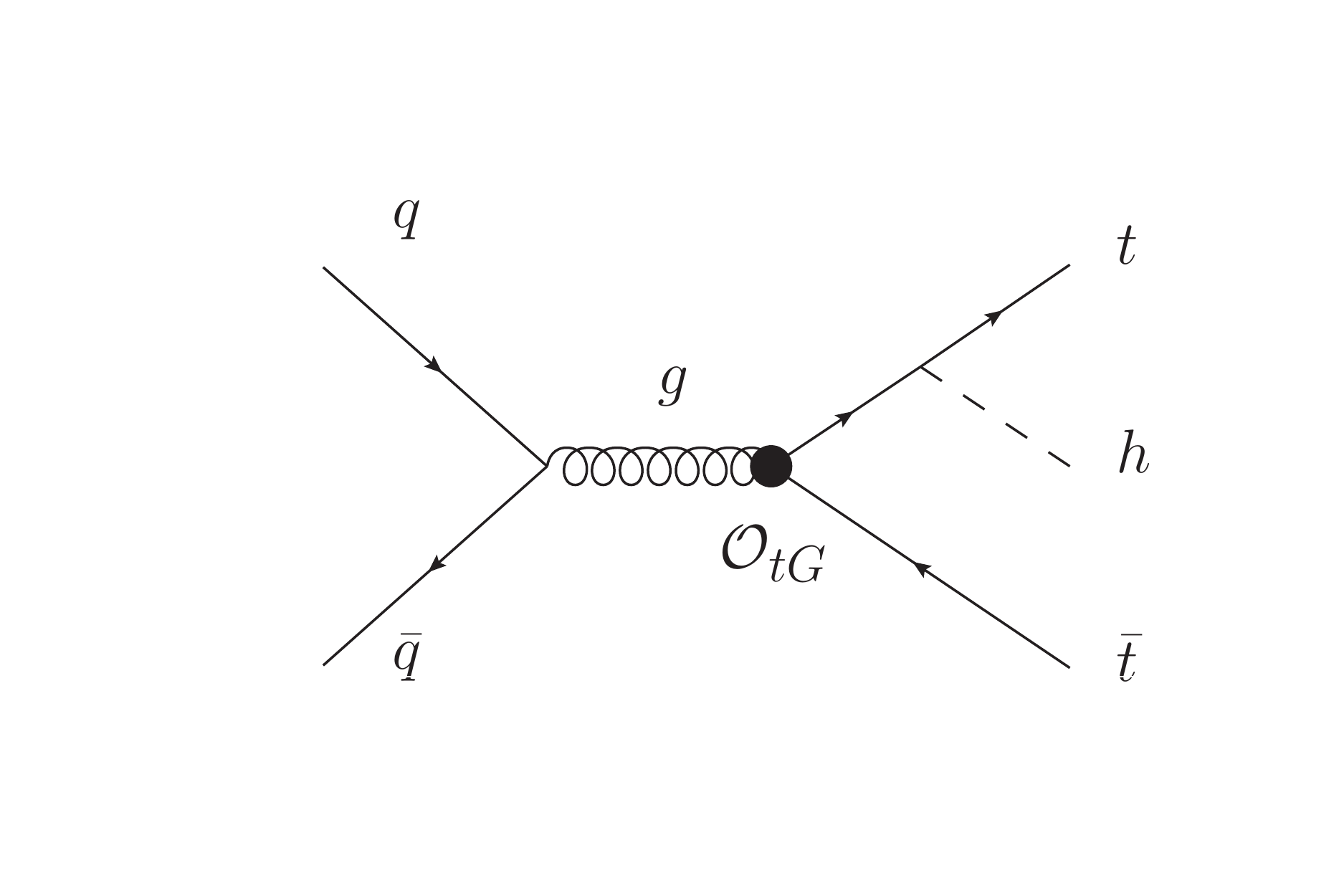}
    } 
 \makebox[\linewidth][c]{%
    \centering
    \includegraphics[width=0.23\textwidth]{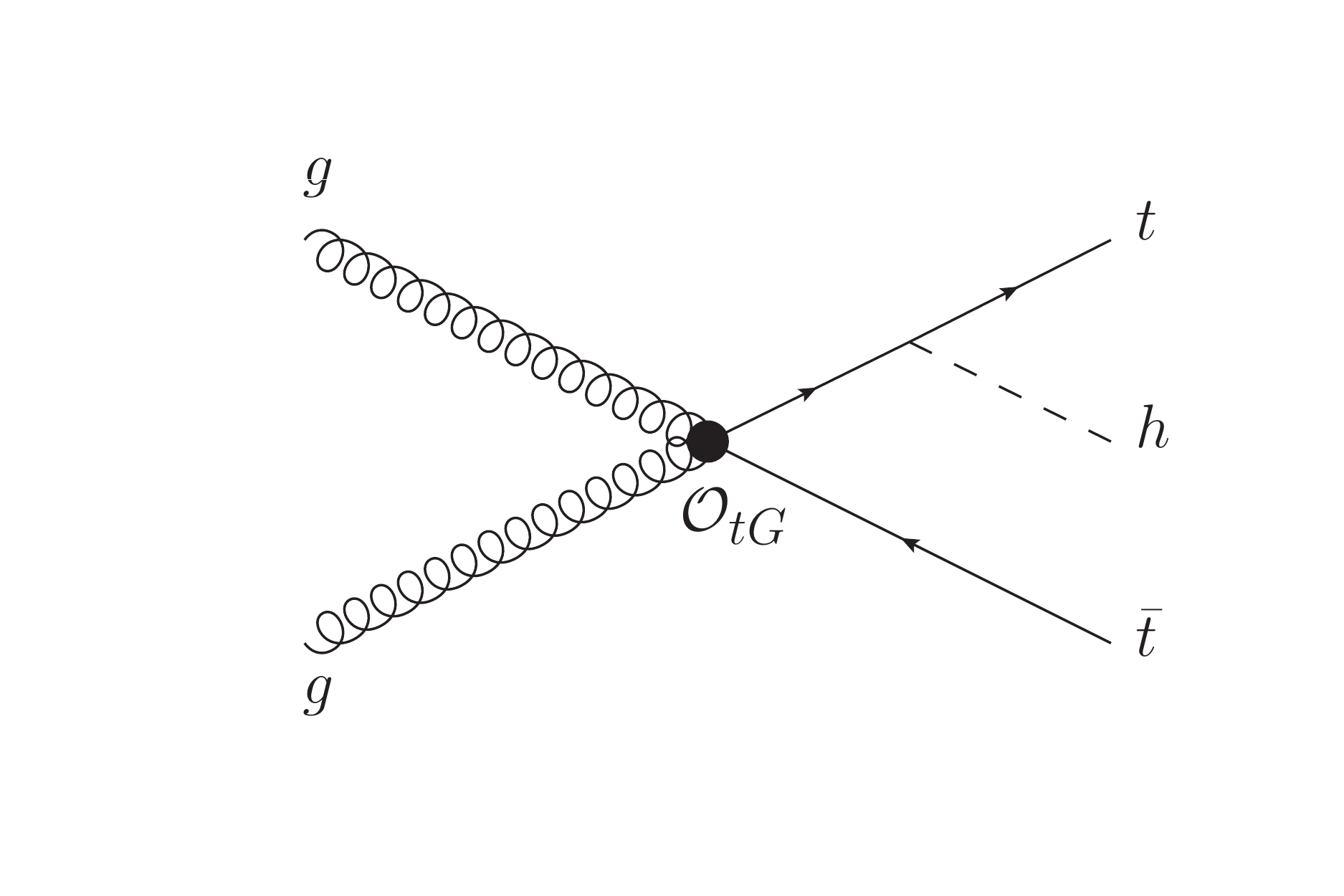}\hspace{-1.3cm}
    \includegraphics[width=0.23\textwidth]{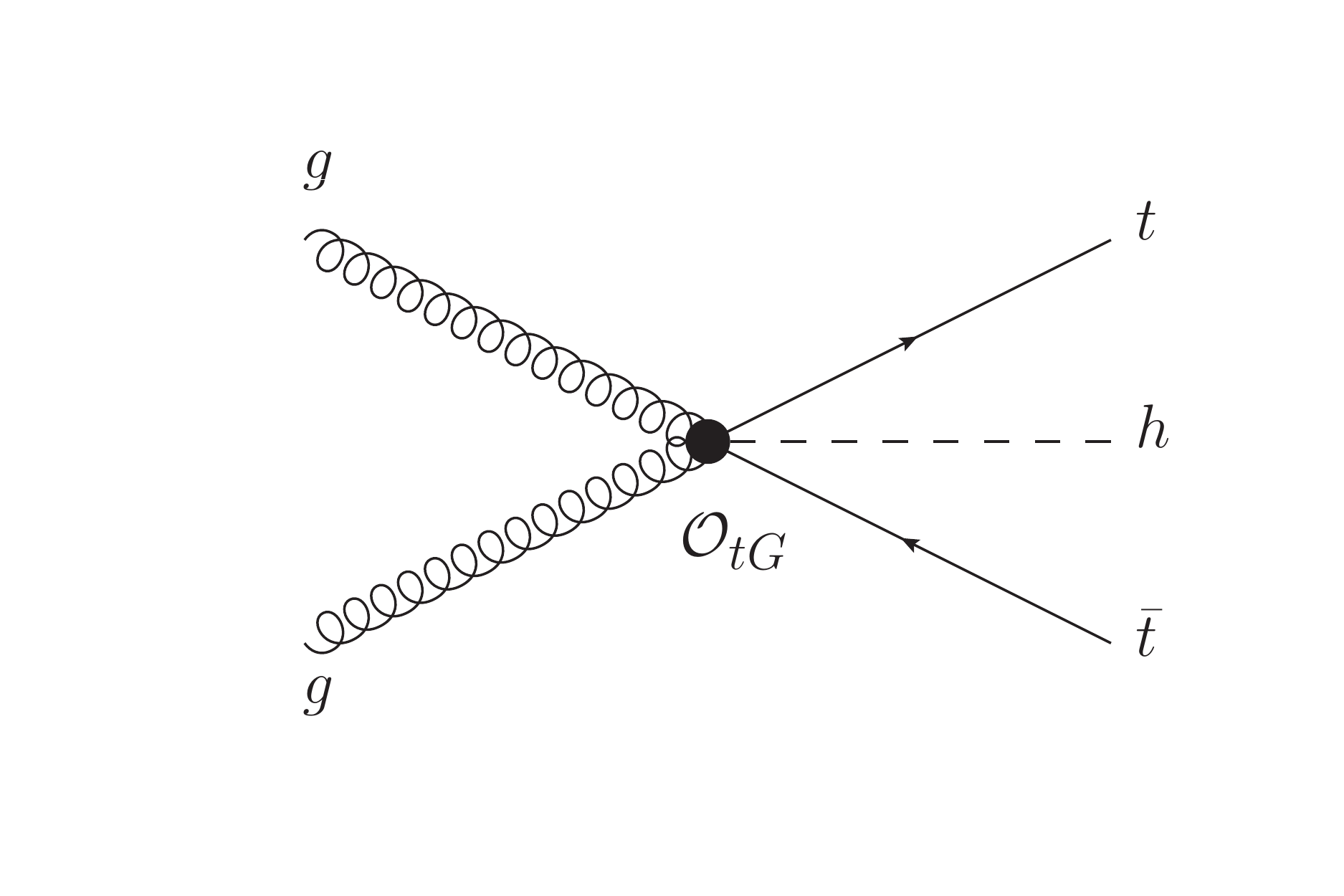}\hspace{-1.3cm}
     \includegraphics[width=0.23\textwidth]{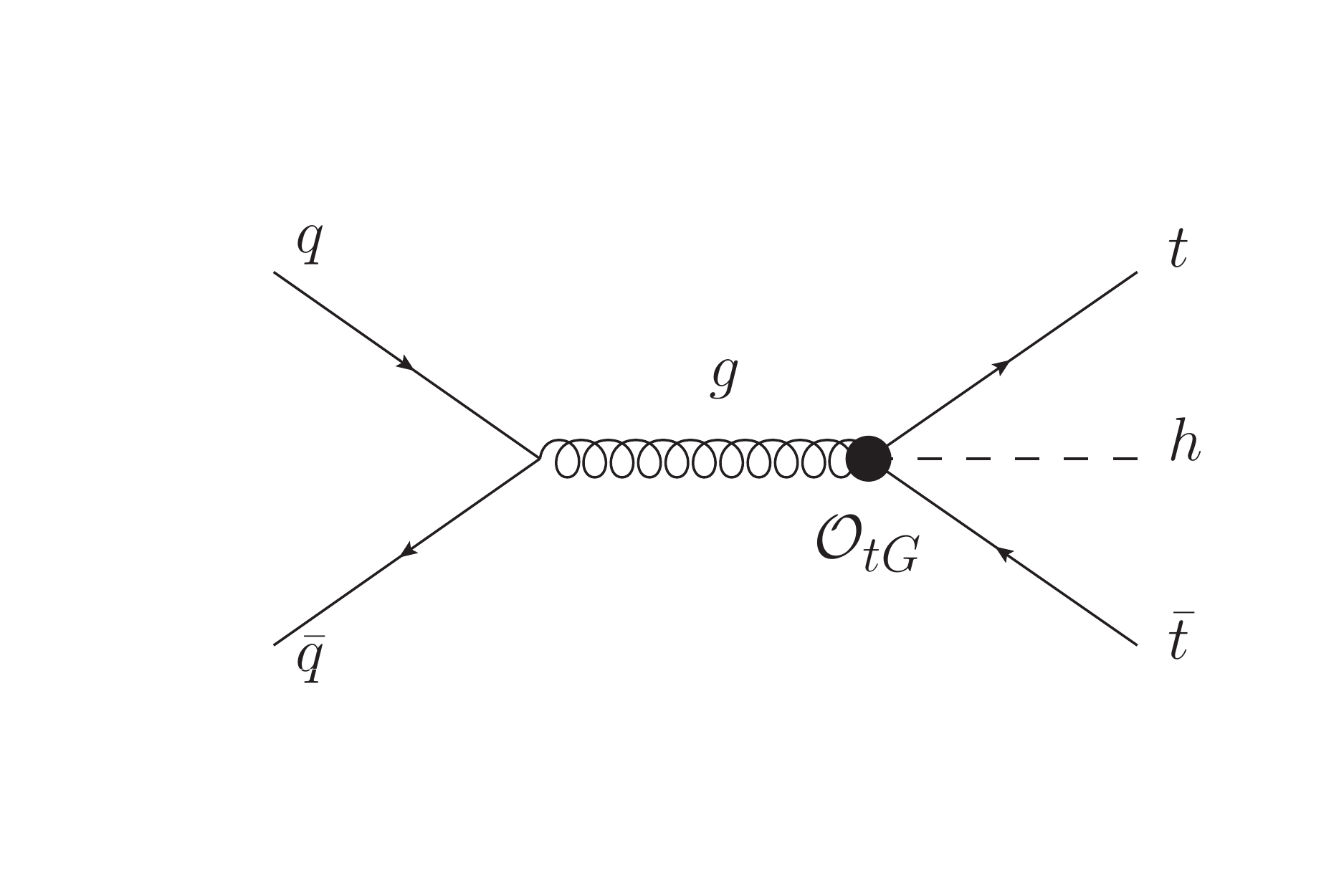} \vspace{-0.5cm}
    }
    \caption{Representative Feynman diagrams contributing to $t \bar{t} h$ production. The black dots represent the BSM vertices arising from the EFT operators.}
    \label{fig:feynman_diagrams}
\end{figure}

We follow the SMEFT framework to study the new physics effects to the Higgs-top coupling at high scales. We adopt the Warsaw basis of operators~\cite{Grzadkowski:2010es}  and focus on two-fermion operators, leading to contributions to $t\bar{t}h$ production at the LHC which are relatively unconstrained
\begin{align}
    \mathcal{O}_{t\phi} & =(H^\dagger H)(\bar{Q}t)\tilde{H}+\text{h.c.}\,, \\
    \mathcal{O}_{tG} & =g_s(\bar{Q}\sigma^{\mu\nu}T_A t)\tilde{H}G_{\mu\nu}^A+\text{h.c.}\,.
\end{align}
The first new physics operator, $ \mathcal{O}_{t\phi}$, rescales the SM top Yukawa coupling. The second one, $\mathcal{O}_{tG}$, corresponds to the chromomagnetic dipole moment of the top-quark. Besides modifying the $gtt$ vertex in the SM, $\mathcal{O}_{tG}$ also gives rise to new interaction vertices, namely $ggtt$, $gtth$ and $ggtth$. While  $\mathcal{O}_{tG}$  results in phenomenological effects to the associated $t\bar{t}$ processes, it amounts  to possibly significant new physics sensitivity in the $t\bar{t}h$ channel \cite{Maltoni:2016yxb}. Hence, we incorporate it in our analysis  exploring its high energy behavior. In Fig.~\ref{fig:feynman_diagrams}, we present a representative set of Feynman diagrams for $t\bar{t}h$ production arising from the  EFT interactions. The experimental LHC analyses constrain these Wilson coefficients at 95\% Confidence Level (CL) to the ranges~\cite{ATLAS-CONF-2020-027,CMS:2018jcg}
\begin{equation}
c_{t\phi}/\Lambda^2=[-2.3,3.1]/\text{TeV}^2,
\ \  c_{tG}/\Lambda^2=[-0.24,0.07]/\text{TeV}^2.
\nonumber 
\end{equation}
Guided by these results, we choose  illustrative values of the coefficients as
\begin{equation}
|c_{tG}/\Lambda^2| =   0.1~\text{TeV}^{-2}\quad {\rm and}\quad  |c_{t\phi}/\Lambda^2| =  1~\text{TeV}^{-2}, 
\label{eq:values}
\end{equation}
for our following representative kinematic distributions. For recent phenomenological SMEFT global fit studies, see Refs.~\cite{Ellis:2020unq,Ethier:2021bye}.

\begin{figure*}[th!]
    \centering
    \includegraphics[width=0.45\textwidth]{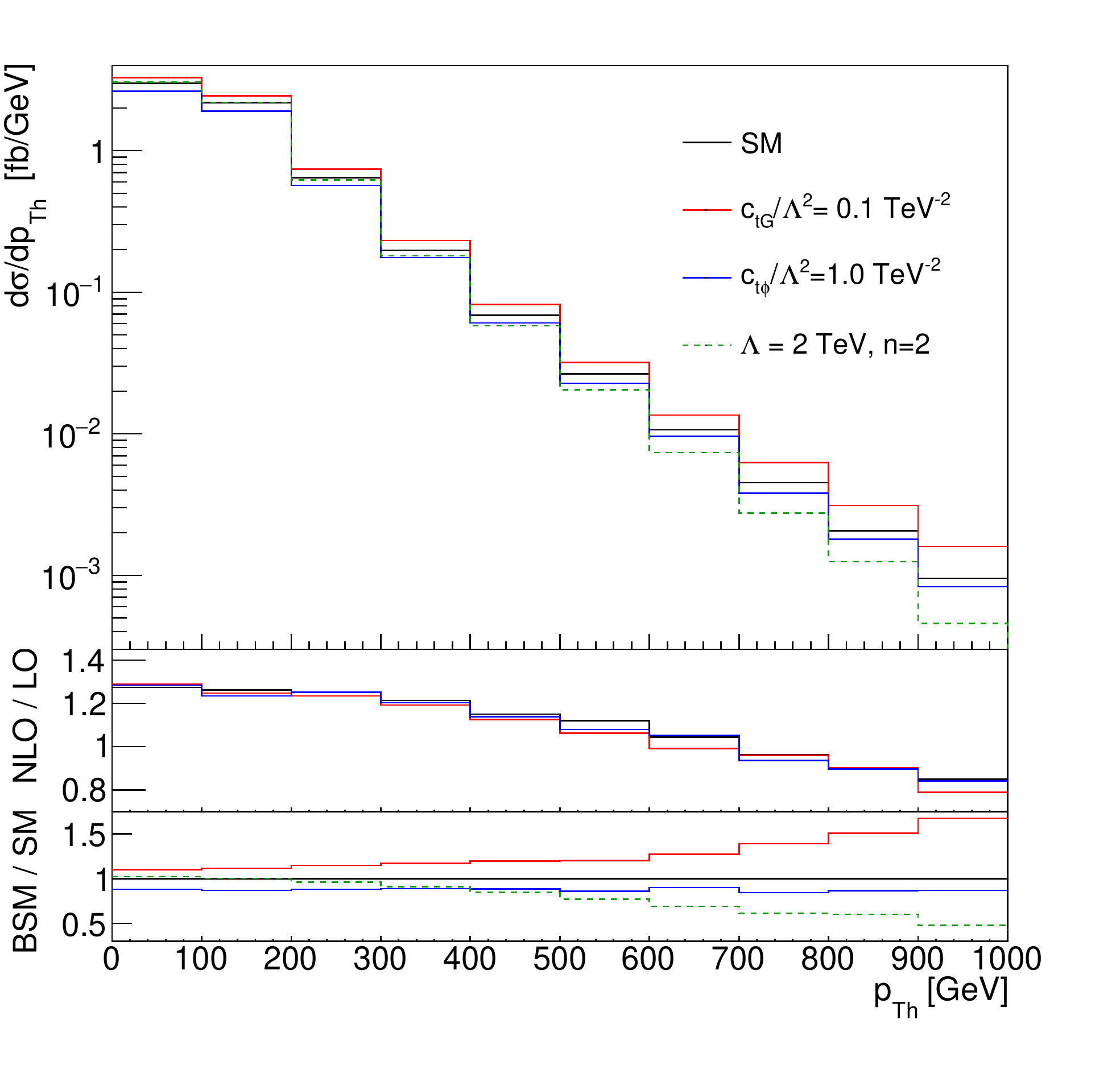} \hspace{0.5cm}
    \includegraphics[width=0.45\textwidth]{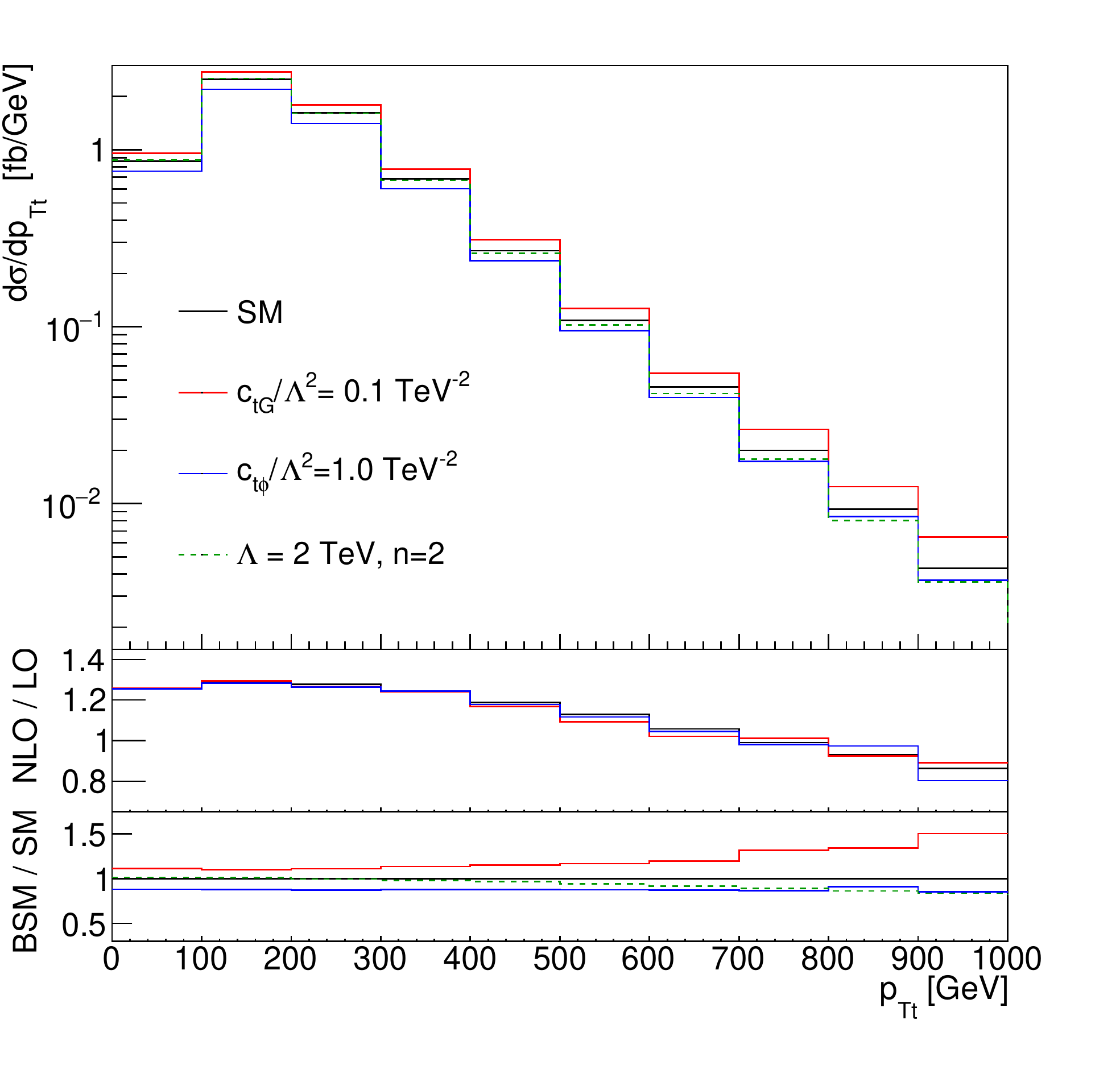}\\
    \includegraphics[width=0.45\textwidth]{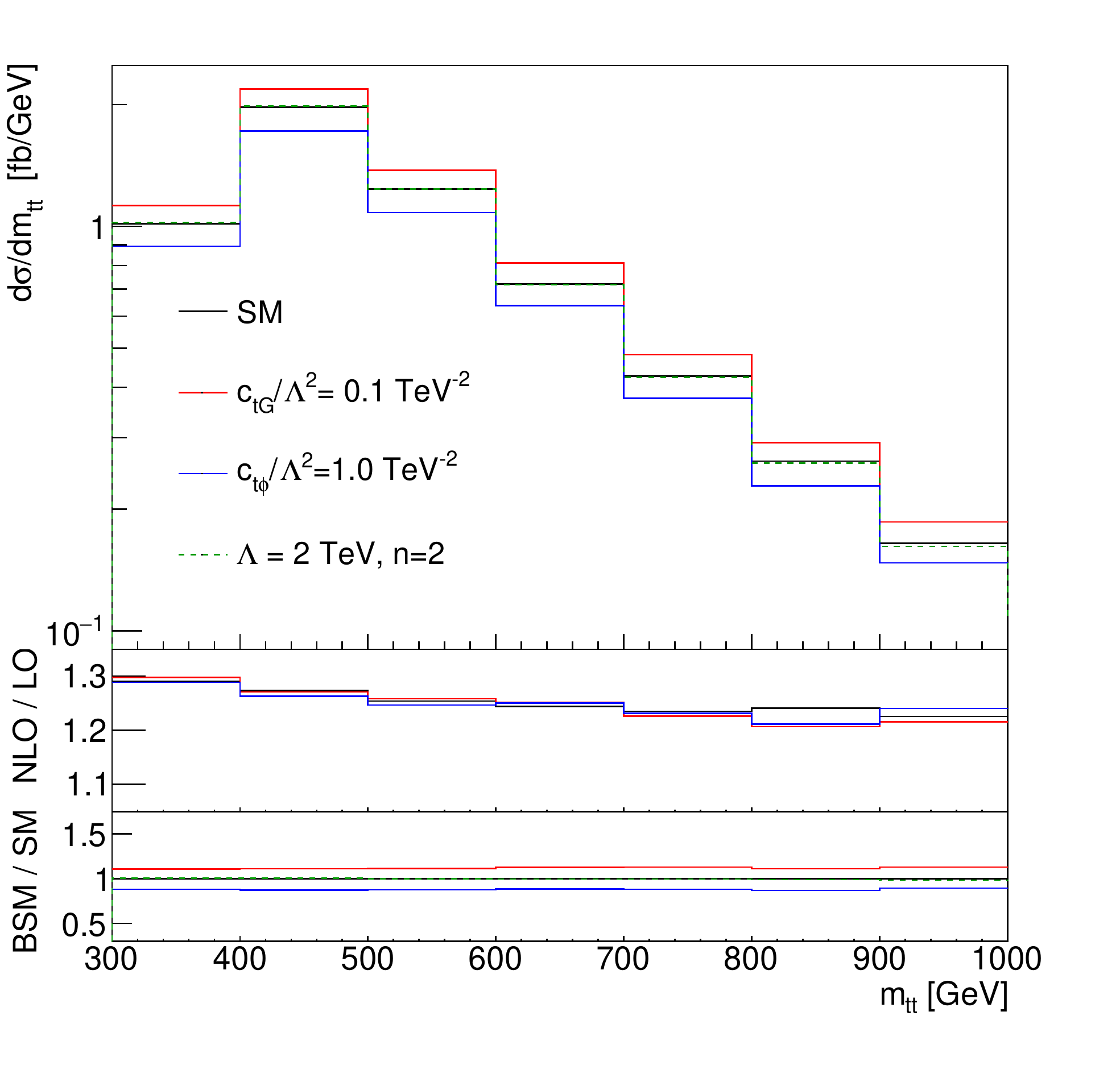}\hspace{0.5cm}
    \includegraphics[width=0.45\textwidth]{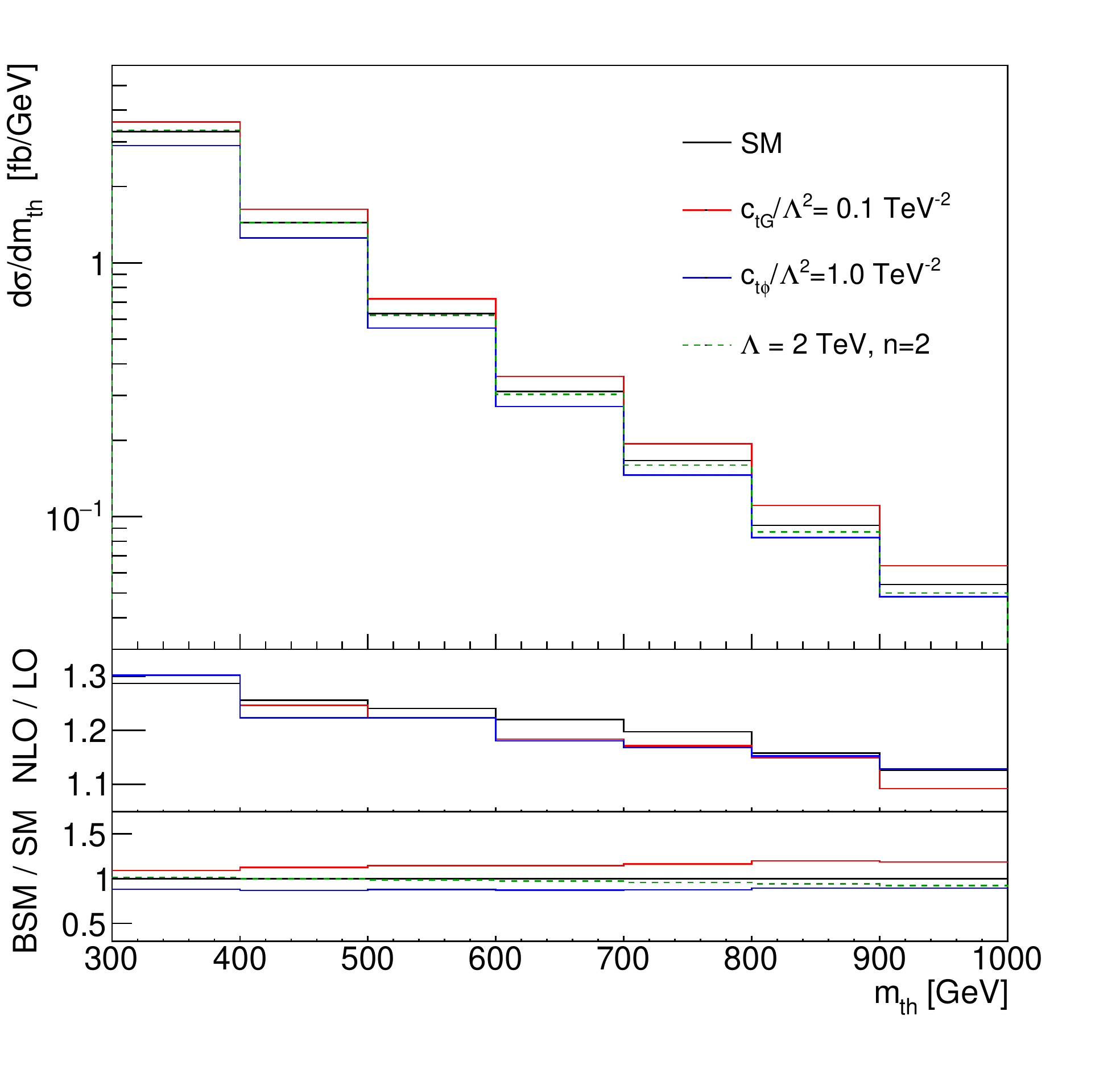}
    \caption{Top panels: Transverse momentum distributions for the Higgs boson $p_{Th}$ (left) and the hardest top-quark $p_{Tt}$ (right). Bottom panels: Invariant mass distributions for the top pair $m_{tt}$ (left) and the Higgs and  top-quark $m_{th}$ (right). Each panel shows on the top the $t\bar{t}h$ sample in the SM and  new physics scenarios.  The results are presented at the NLO QCD fixed order.  We also show the local NLO $K$-factor (middle panel in each figure as NLO/LO) and the ratio between new physics and SM scenarios (bottom panel in each figure as BSM/SM). We assume the LHC at 14~TeV.} 
    \label{fig:pth_parton_level}
\end{figure*}

\subsection{Higgs-Top coupling form-factor}
\label{sec:FF}

The top-quark Yukawa coupling has a special role in the naturalness problem, displaying the  dominant quantum corrections to the Higgs mass. Thus, it is crucially important to probe the Higgs-top interaction at high scales into the ultra-violet regime.
It is well-motivated to consider that the top-quark and Higgs boson may not be fundamental, but composite particles arising from strongly interacting new dynamics at a scale $\Lambda$~\cite{Pomarol:2012qf,Panico:2015jxa,Liu:2017dsz,Banerjee:2021qhr}.  In such scenarios, 
the top Yukawa may exhibit a momentum-dependent form-factor near or above the new physics scale $\Lambda$, rather than a point-like interaction. It is challenging to write a form-factor, in a general form, without prior knowledge of the underlying strong dynamics of the specific composite scenario. Inspired by the nucleon form-factor~\cite{Punjabi:2015bba}, we adopt the following phenomenological ansatz
\begin{align}
    \Gamma(Q^2/\Lambda^2)=\frac{1}{\left(1+Q^2/\Lambda^2\right)^n}\,,
\end{align}
where $Q$ is the energy scale associated with the physical process. This educated guess results in a dipole form-factor for the $n=2$ scenario with an exponential spatial distribution in a space-like probe. Higher values of $n$ correspond to higher multi-poles, typically leading to a stronger suppression.

\section{Analysis}
\label{sec:LHC}

\begin{figure}[b]
    \centering
    \includegraphics[width=0.5\textwidth]{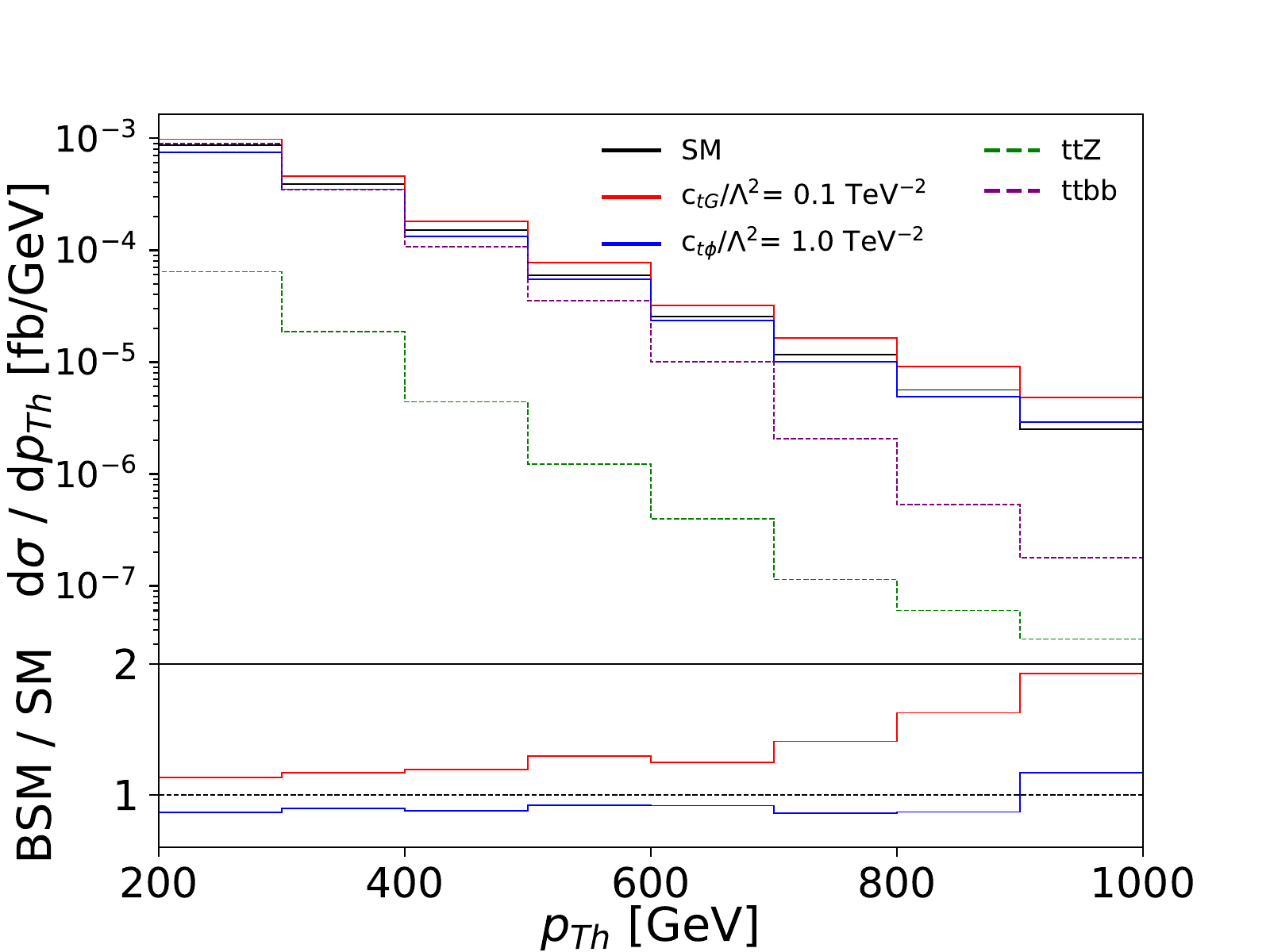}
    \caption{Transverse momentum distribution of the Higgs boson $p_{Th}$ for the  $t\bar{t}h$ sample in the SM (black) and new physics scenarios with $c_{tG}/\Lambda^2=0.1~\text{TeV}^{-2}$ (red), $c_{t\phi}/\Lambda^2=1~\text{TeV}^{-2}$ (blue). The leading backgrounds $t\bar{t}b\bar{b}$ (purple) and $t\bar{t}Z$ (green)  are also presented.  We assume the LHC at 14~TeV.}
    \label{fig:pth_eft}
\end{figure}

\begin{figure}[t]
    \centering
    \includegraphics[width=0.5\textwidth]{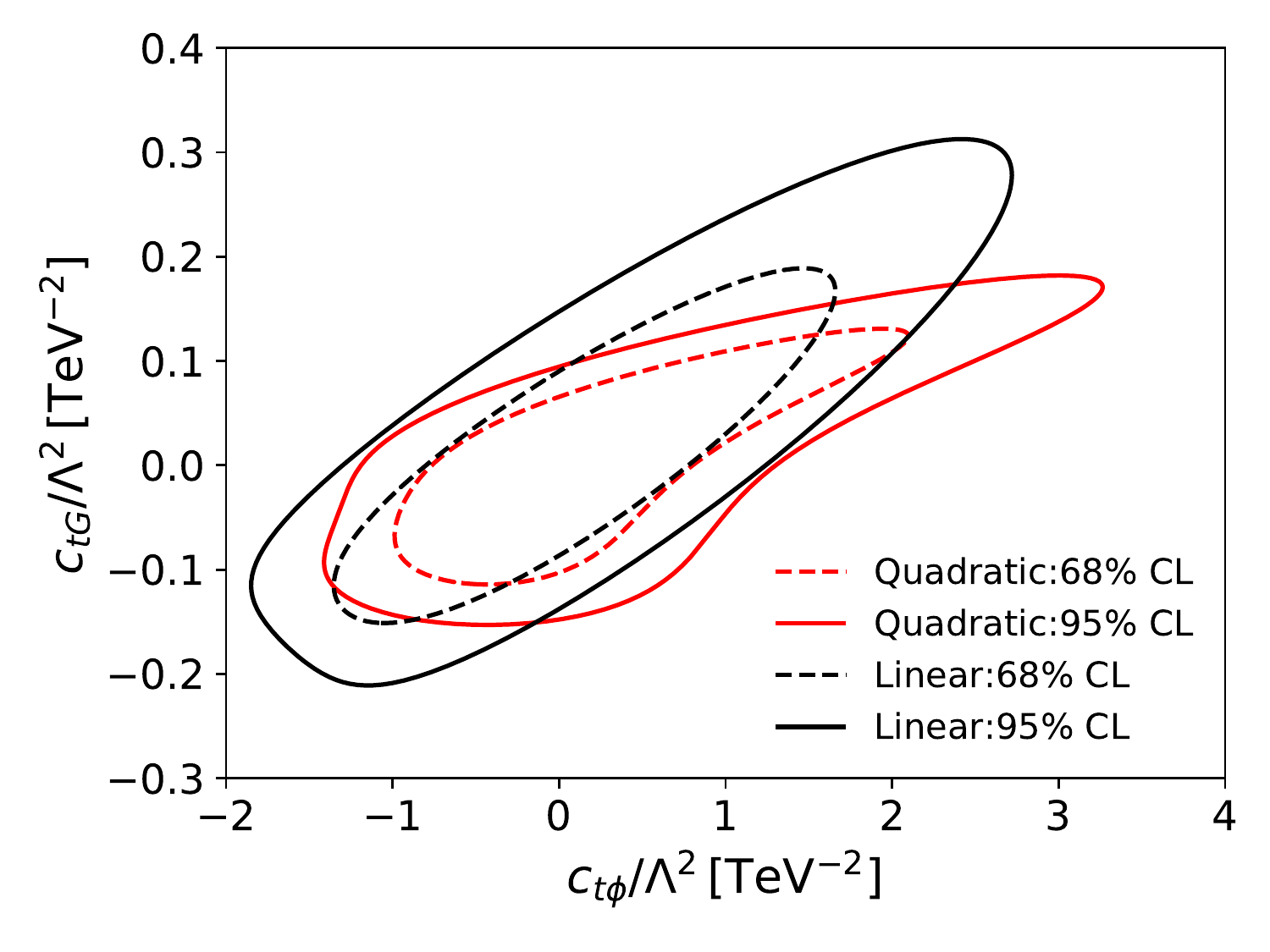}
    \caption{95\% (full line) and 68\% (dashed line) CL  limits on the Wilson coefficients $(c_{tG}/\Lambda^2,c_{t\phi}/\Lambda^2)$ at the 14~TeV HL-LHC with 3~$ab^{-1}$ of data. The results are presented both at the linear (black) and  quadratic (red) order in dimension-6 SMEFT operator coefficients.
    }
    \label{fig:limit_eft}
\end{figure}

To probe these new physics contributions, we explore the ${pp\to t\bar{t}h}$ channel at high energy scales. We combine the large signal event rate with controlled backgrounds, studying the boosted $h\rightarrow b\bar{b}$ final state in association with leptonic top-quark pair decays. The signal is defined in the four $b$-tag sample  and displays two opposite sign leptons. The leading backgrounds, in order of relevance, are $t\bar{t}b\bar{b}$ and $t\bar{t}Z$. 

We perform the signal and background event generation with {\tt MadGraph5\_aMC@NLO}~\cite{Alwall:2014hca}. The $t\bar{t}h$ and $t\bar{t}Z$ samples are generated at NLO QCD and the $t\bar{t}b\bar{b}$ sample at LO. The dimension-six EFT contributions are added through the FeynRules model {\tt SMEFT@NLO}~\cite{Degrande:2020evl}. This implementation grants one-loop QCD computations, accounting for the EFT contributions. In particular, it incorporates  relevant extra radiation effects at the matrix element level~\cite{Goldouzian:2020ekx}.  Shower, hadronization, and underlying event effects are simulated with {\tt Pythia8}~\cite{Sjostrand:2014zea} using the {\tt Monash} tune~\cite{Skands:2014pea}. We use {\tt MadSpin} to properly describe the top-quark decays, accounting for spin correlation effects~\cite{Artoisenet:2012st}. We adopt the parton distribution functions from MMHT2014 NLO with $\alpha_S(m_Z)=0.118$~\cite{Harland-Lang:2014zoa} in the five flavor scheme. Additional relevant parameters are $m_t=172$~GeV, $m_h=125$~GeV, $m_Z=91.1876$~GeV, $m_W=79.82$~GeV, and $G_F=1.16637\times 10^{-5}\text{~GeV}^{-2}$. We set our scales to a constant value of $\mu_F=\mu_R=m_t+m_h/2$ to align better with previous studies~\cite{Maltoni:2016yxb}. We assume the LHC at $\sqrt{s}=14$~TeV.

Robust new physics studies at the LHC usually  come hand in hand with precise theoretical calculations. The impact of the  higher order QCD corrections, which  can be conventionally estimated by a $K$-factor (\emph{i.e.} the  ratio between the NLO and LO predictions), usually result in significant contributions. To illustrate the higher order and new physics effects at high energies, we present in Fig.~\ref{fig:pth_parton_level} the NLO fixed order parton level distribution for several relevant kinematic observables associated with the $t\bar{t}h$ signal sample:
the transverse momentum distribution for the Higgs boson $p_{Th}$ (upper left), for the hardest top-quark $p_{Tt}$ (upper right), the invariant mass distribution for the top pair $m_{tt}$ (lower left), and for the Higgs and  top-quark $m_{th}$ (lower right). 
We observe that the higher order QCD corrections are correlated with the kinematic observables, resulting in about $20\%-30\%$ variation (as seen in the panels of NLO/LO) and cannot be captured by a global NLO $K$-factor. It is thus crucial to include the higher order  predictions in the full differential analysis. 

New physics contributions may sensitively depend on the kinematics as well, as demonstrated in the panels of BSM/SM in Fig.~\ref{fig:pth_parton_level}. 
High transverse momenta of an on-shell top quark or Higgs boson could probe the space-like regime for the top-Higgs interactions, while the high invariant mass of the $tH$ system could be sensitive to the time-like regime from heavy states in $s$-channels.
First, we observe sizable energy enhancement arising from the $\mathcal{O}_{tG}$ operator, in particular, for the transverse Higgs momentum distribution (as seen in the panels of  BSM/SM), starting with a 10\%  increase at the non-boosted regime $p_{Th}<100$~GeV, adding up to 65\% for $p_{Th}=1$~TeV. In contrast, due to the generic dipole suppression, the form-factor scenario displays a depletion in cross-section at higher energies. The rate is reduced by $5\%$ at $p_{Th}=200$~GeV, reaching 55\% suppression at $p_{Th}=1$~TeV. 
For the form-factor scenario, we adopt a representative scale $Q=p_{Th}$.
New physics effects associated with the operator $\mathcal{O}_{t\phi}$ do not result in a distinct energy profile with respect to the SM. In the $t\bar{t}h$ process, this operator only contributes with a  shift to the top Yukawa, resulting in a flat rescale with respect to the SM cross-section, independent of the process energy scale. Despite the absence of a manifest energy enhancement, this new physics contribution can also benefit from our high energy scale analysis due to more controlled backgrounds at the boosted Higgs regime, as we will show in the following.

The boosted Higgs analysis, in combination with jet substructure techniques  effectively suppress the initially  overwhelming backgrounds for the  $t \bar t h$ signal with the dileptonic top decays and $h\to b\bar{b}$, as first shown in Ref.~\cite{Buckley:2015vsa}. Here we follow a similar strategy. We start our analysis requiring two isolated and opposite sign leptons with $p_{T\ell} >10$~GeV and $|\eta_\ell|<3$. For the hadronic component of the event, we first reconstruct jets with the Cambridge-Aachen algorithm with $R=1.2$~\cite{Cacciari:2011ma},  requiring at least one boosted fat-jet with $p_{TJ}>200$~GeV and $|\eta_J|<3$. We demand that one of the fat-jets be Higgs tagged with the Butterworth-Davison-Rubin-Salam (BDRS) algorithm~\cite{Butterworth:2008iy,Plehn:2009rk}. Higgs tagging of the fat-jet via the BDRS algorithm involves identifying three subjets within the fat-jet. This is done by shrinking the jet radius until the fat-jet splits into three filtered jets. The radius of separation among the filtered jets is defined as $R_\text{filt}=\text{min}(0.3,R_{bb}/2)$. Among the three filtered jets, the two hardest are required to be $b$-tagged, while the third filtered jet tracks the dominant $\mathcal{O}(\alpha_s)$ radiation from the Higgs decay. 

As we only have one hadronic heavy particle decay, namely the Higgs boson, we proceed with the event reconstruction using a smaller jet size to further reduce the  underlying event contamination. Thus, we remove all the hadronic activity associated with the Higgs fat-jet and re-cluster the remaining particles with the jet radius $R=0.4$, using the anti-k$_t$ jet algorithm. We demand two $b$-tagged jets with $p_{tb}>30$~GeV and $|\eta_b|<3$. As our final state displays in total four $b$-tagged jets, we exploit the  improvements in the central tracking system, that will be in operation for the HL-LHC run, to enhance the event rate for our signal. Based on the ATLAS report~\cite{CERN-LHCC-2017-021}, we assume 85\% $b$-tagging efficiency and 1\% mistag rate for light-jets. To further suppress the backgrounds, the filtered  mass for the Higgs candidate is imposed to be around the Higgs boson mass $|m_h^\text{BDRS}-125\ {\rm GeV}|<10$~GeV. We show in Table~\ref{tab:cut_flow} more details on the cut-flow analysis. 

\begin{table}[h!]
\centering
\begin{tabular}{l | c  | c | c  }
  \multicolumn{1}{c|}{cuts} &
  \multicolumn{1}{c|}{$t\bar{t}h$}&
  \multicolumn{1}{c|}{$t\bar{t}b\bar{b}$} &
  \multicolumn{1}{c}{$t\bar{t}Z$}
   \\
  \hline
BDRS $h$-tag, $p_{T\ell}>10$~GeV,  $|\eta_\ell|<3$, $n_\ell= 2$ & 3.32 &  6.35 & 1.02 \\ 
$p_{Tj}>30$~GeV,  $|\eta_j|<3$, $n_j\ge 2$, $n_b$=2 & 0.72 &  1.97 & 0.22 \\ 
$|m_{h}^{\text{BDRS}}-125|<10$~GeV  & 0.15 & 0.14 & 0.009 \\ 
\end{tabular} 
\caption{Cut-flow for signal and backgrounds at LHC ${\sqrt{s}=14~\text{TeV}}$. The selection follows the BDRS analysis described in the text. Rates are in units of fb and account for  85\% (1\%) $b$-tag (mistag) rate, hadronization, and underlying event effects. }
\label{tab:cut_flow}
\end{table}

\subsection{Scale for the EFT operators}
\label{sec:ReachEFT}

In Fig.~\ref{fig:pth_eft}, we go beyond the partonic level calculation and display the hadron level transverse momentum distribution ($p_{Th}$) for the Higgs boson candidate from the $pp\to t\bar{t}h$ channel in the SM and the EFT contributions, in addition to the leading backgrounds $t\bar{t}b\bar{b}$ and $t\bar{t}Z$. We observe that the boosted Higgs search dovetails nicely with our BSM physics study as presented in Fig.~\ref{fig:pth_parton_level}. At the higher energy scales, both the backgrounds get further depleted and the new physics effects become more prominent. In particular, we observe a large enhancement from the $\mathcal{O}_{tG}$ contributions at the high energy scales.

\begin{figure}[b!]
    \centering
    \includegraphics[width=0.5\textwidth]{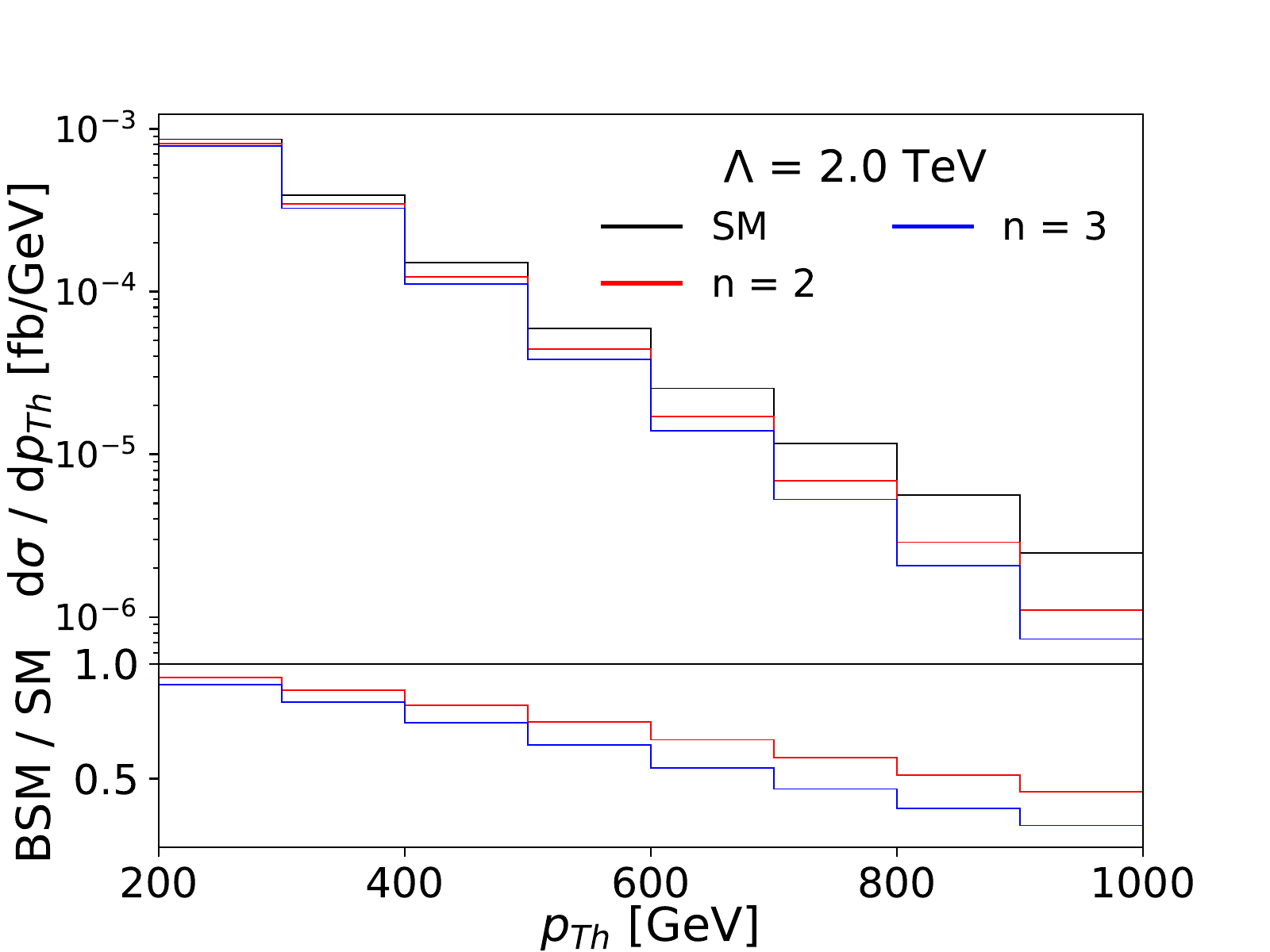}
        \caption{Transverse momentum distribution of the Higgs boson $p_{Th}$ for the  $t\bar{t}h$ sample in the SM (black) and new physics scenarios with $n=2$ (red) and $n=3$ (blue), assuming $\Lambda=2$~TeV. We assume the LHC at 14~TeV.}
    \label{fig:pth_form}
\end{figure}

To explore the sensitivity reach for these effects in the boosted regime, we perform a binned log-likelihood analysis on the $p_{Th}$ distribution. In Fig.~\ref{fig:limit_eft}, we present the 68\% and 95\% CL limits on the Wilson coefficients $(c_{tG}/\Lambda^2,c_{t\phi}/\Lambda^2)$. We assume the HL-LHC at 14~TeV with 3~ab$^{-1}$ of data. To infer the uncertainty on the EFT expansion, we present the results accounting for terms up to linear and quadratic order on the Wilson coefficient  $c_i/\Lambda^2$. We observe only small differences between these two scenarios,  which is a good indication of the robustness of our results. 

CMS has recently reported an EFT interpretation using associated top quark production data with an integrated luminosity of $\mathcal{L}=41.5~\text{fb}^{-1}$~\cite{CMS-PAS-TOP-19-001}. The signal samples include, in  particular, the $t\bar{t}h$ and $thq$ processes, being direct sensitive to the top-quark Yukawa coupling. The resulting constraint at the 95\% CL for the chromomagnetic operator leads to two regions $c_{tG}/\Lambda^2=[-1.26, -0.69]~\text{TeV}^{-2}$ and $[0.08, 0.79]~\text{TeV}^{-2}$. The same holds for the $\mathcal{O}_{t\phi}$ operator where $c_{t\phi}=[-14.12, -1.46]~\text{TeV}^{-2}$ and $[32.30, 44.48]~\text{TeV}^{-2}$. While CMS does not focus on the very high energy scales and uses the leptonic Higgs decays, we explore the largest Higgs branching ratio, $h\to b\bar{b}$, in the boosted Higgs regime, and thus obtaining  significantly higher sensitivities at the HL-LHC.

\subsection{Probing the form-factor}
\label{sec:ReachFF}

\begin{figure}[t!]
    \centering
    \includegraphics[width=0.5\textwidth]{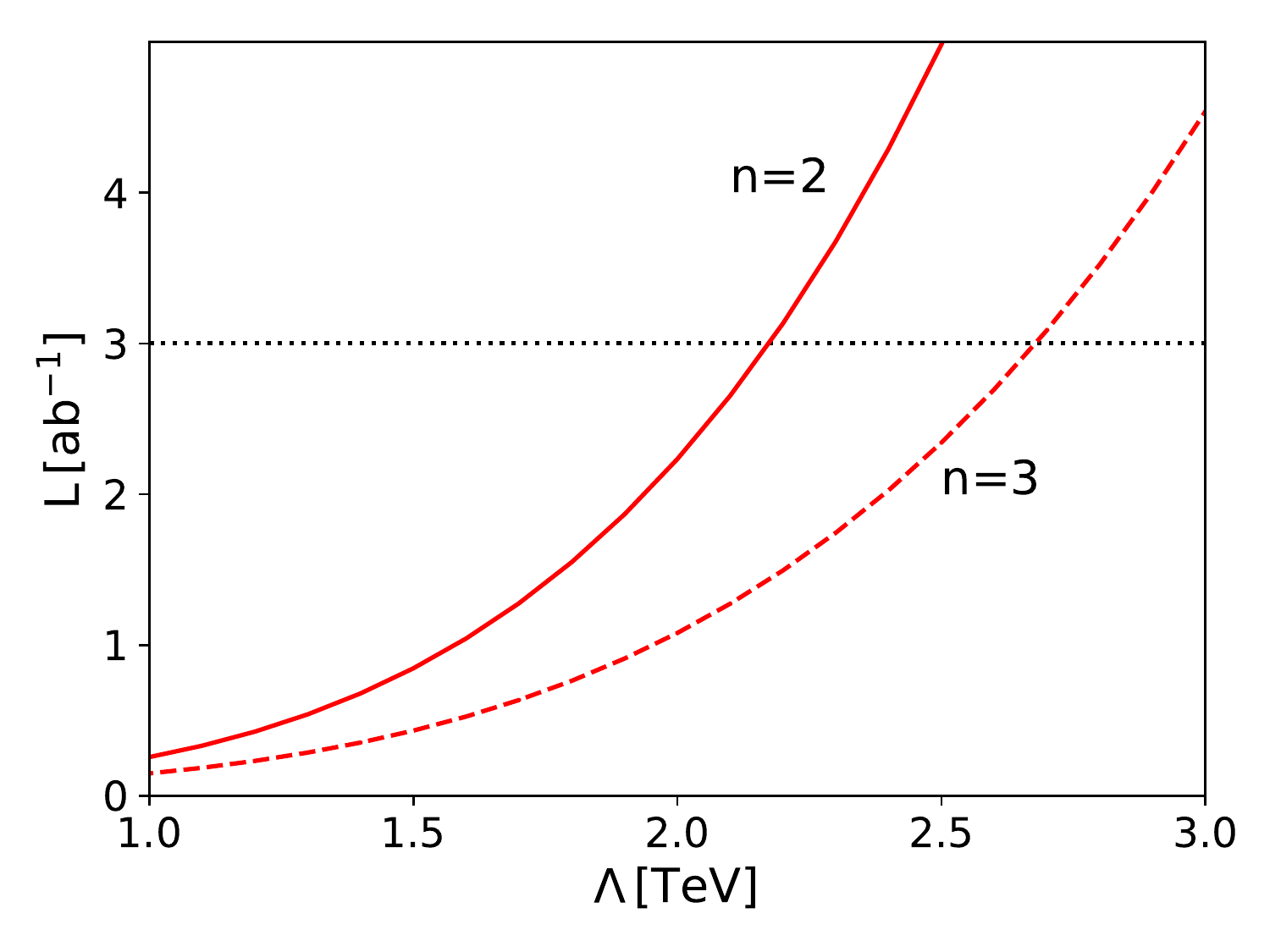}
    \caption{95\% CL sensitivity on the new physics scale $\Lambda$ as a function of the LHC luminosity. We consider two form-factor scenarios: $n=2$ (solid line) and $n=3$ (dashed line).}
    \label{fig:limit_ff}
\end{figure}

\begin{table*}[t!]
\centering
\begin{tabular}{|c|c|c|c|}
\hline
\multirow{2}{*}{} & \multirow{2}{*}{channel} & $c_i/\Lambda^2~[\text{TeV}^{-2}]$    & $\Lambda/\sqrt{c_i}$~[TeV]   \\
                  &                          &  95\% CL bounds                      &   BSM scale   \\
\hline
\multirow{4}{*}{$c_{t\phi}$} & $t\bar{t}h$  (this work)                                           &  [$-1.04$, 1.00]    &  1.0  \\
                   & $h^* \to ZZ \to\ell\ell\nu\nu$~\cite{Goncalves:2020vyn}    &  [$-2.8$ , 1.5]     &  0.6  \\
                   & $h^* \to ZZ \to 4\ell$~\cite{Goncalves:2018pkt}            &  [$-3.3$ , 3.3]     &  0.55  \\
                     &Higgs comb. ATLAS~\cite{ATLAS-CONF-2020-027}                   &  [$-2.3$ , 3.1]      &  0.57  \\
\hline
\multirow{2}{*}{$c_{tG}$} &  $t\bar{t}h$    (this work)  &  $[-0.11$ , 0.12] &   2.9    \\
                          & $t\bar{t}$ CMS~\cite{CMS:2018jcg}       &  [$-0.24$ , 0.07] &   2.1   \\
\hline
\multirow{3}{*}{form-factor $n=2$} & $t\bar{t}h$             (this work)                                       &  -   &   2.1  \\
                                   & $h^* \to ZZ \to\ell\ell\nu\nu$~\cite{Goncalves:2020vyn}    &  -   &   1.5  \\
                                   & $h^* \to ZZ \to 4\ell$~\cite{Goncalves:2018pkt}            &  -   &   0.8  \\
\hline                    
\multirow{3}{*}{form-factor $n=3$} & $t\bar{t}h$                                    (this work)    &  -  &   2.7  \\
                                   & $h^* \to ZZ \to\ell\ell\nu\nu$~\cite{Goncalves:2020vyn}    &  -   &   2.1  \\
                                   & $h^* \to ZZ \to 4\ell$~\cite{Goncalves:2018pkt}            &  -  &   1.1  \\
\hline
\end{tabular}
\caption{Summary results from the $t\bar t h$ studies for the Higgs-top coupling at high scales in terms of the dimension-6 operators and general form-factor scenarios. The results are shown at 95\% CL, and we assume the HL-LHC at 14~TeV with 3 ab$^{-1}$ of data. For comparison, we also show the results from off-shell $h^*$ studies, the ATLAS Higgs combination with $139~\text{fb}^{-1}$, and the CMS top pair bound with $35.9~\text{fb}^{-1}$.}
\label{tab:summary}
\end{table*}

In Fig.~\ref{fig:pth_form}, we present the transverse momentum distribution ($p_{Th}$) for the Higgs boson candidate from the $pp\to t\bar{t}h$ channel in the SM and the form-factor  contribution. We consider two  hypotheses $n=2$ and $n=3$ with the new physics scale $\Lambda=2$~TeV. While it is challenging to probe the BSM effects at relatively small scales, these contributions can be effectively enhanced at the boosted regime. For instance, starting at $p_{Th}\sim 200$~GeV with $n=2$ ($n=3$), we observe a 5\% (9\%) effect. Moving to $p_{Th}\sim 400$~GeV, the new physics results in larger depletion of 18\% (25\%) with respect to the SM hypothesis.

Our relatively large event rate with the boosted $h\to b\bar{b}$ analysis, grants probes at large energy scales with relevant statistics. Hence, we explore the full profile of the $p_{Th}$ distribution through a binned log-likelihood analysis. The new physics sensitivity is presented in Fig.~\ref{fig:limit_ff}. The HL-LHC, with 3 ab$^{-1}$ of data, will be able to probe these new physics effects up to a scale of $\Lambda=2.1$~TeV for $n=2$ and $\Lambda=2.7$~TeV for $n=3$ at 95\% CL. These results are complementary to the off-shell Higgs analyses, $gg\to h^{*}\to ZZ$. For the latter, assuming $n=3$, the limits on the new physics scale are $\Lambda=1.1$~TeV for the $4\ell$ final state and $\Lambda=2.1$~TeV for the $\ell\ell\nu\nu$ final state~\cite{Goncalves:2018pkt,Goncalves:2020vyn}.

\section{Summary and discussions}
\label{sec:Sum}

 We studied the prospects to \emph{directly} probe the Higgs-top coupling for new physics at high energy scales using the $pp\to t\bar{t}h$ process at the HL-LHC. We considered two beyond the SM scenarios, namely the SMEFT framework and a general Higgs-top form-factor, as discussed in Sec.~\ref{sec:BSM}. We presented in Sec.~\ref{sec:LHC} the general phenomenological effects for these new physics  contributions, showing that they could produce augmented new physics effects at high energy scales.
 
 Focusing on the boosted Higgs regime in association with jet substructure techniques, we explored the largest Higgs branching fraction $h\to b\bar{b}$ along with the clean leptonic top-quark decays. The BSM effects were constrained through a shape analysis on the  $p_{Th}$ spectrum. We observed the potential sensitivity at the  TeV-scale for new physics both in the EFT and form-factor scenarios. The chromomagnetic dipole operator was probed up to ${\Lambda/\sqrt{c_{tG}}\approx 2.9}$~TeV and the $\mathcal{O}_{t\phi}$ operator to ${\Lambda/\sqrt{c_{t\phi}}\approx 1.0}$~TeV, as shown in  Sec.~\ref{sec:ReachEFT}. The limits presented  sub-leading differences between the linear and quadratic $c_i/\Lambda^2$ expansion, indicating that our phenomenological study satisfies the EFT expansion. Finally, when considering a more general  Higgs-top quark form-factor in Sec.~\ref{sec:ReachFF}, we concluded that the HL-LHC is sensitive to new physics up to the scale $\Lambda=2.1$~TeV for $n=2$ and $2.7$~TeV for $n=3$ at 95\%~CL. Further details are summarized in Table~\ref{tab:summary}. The $t\bar t h$ studies at high scales, which \emph{directly} explore the Higgs-top Yukawa interaction, results in a competitive and complementary pathway for BSM sensitivity in comparison to the off-shell Higgs channels and the current ATLAS and CMS limits.   

 Some improvements in sensitivity can be anticipated by including other modes, such as $t\bar{t}(h\to \gamma\gamma)$, which would yield a cleaner signal but a lower rate~\cite{Brehmer:2019xox}. In addition, we can increase our present $t\bar{t}(h\to b\bar{b})$  statistical sample by about a factor of six, if we include one leptonic decay plus one hadronic decay of the $t\bar t$. The analysis, however, would be more complex, with significantly larger QCD  backgrounds~\cite{Aaboud:2017rss}. Finally, while we adopt {\tt MadGraph5\_aMC@NLO} as our general Monte Carlo generator (that accounts for the signal EFT contributions at NLO QCD), we acknowledge some other recent important developments associated with the $t\bar{t}b\bar{b}$ background~\cite{Jezo:2018yaf,Denner:2020orv,Bevilacqua:2021cit}. We leave those improvements to future work with realistic simulations. 

\begin{acknowledgments}
RMA and DG thank the U.S.~Department of Energy for the financial support, under grant number DE-SC 0016013. The work of TH, SCIL, and HQ is supported by the U.S.~Department of Energy under grant No.~DE-FG02-95ER40896 and by the PITT PACC. Some computing for this project was performed at the High Performance Computing Center at Oklahoma State University, supported in part through the National Science Foundation grant OAC-1531128.
\end{acknowledgments}

\bibliographystyle{unsrt}
\bibliography{references}

\begin{thebibliography}{10}

\bibitem{Hill:2002ap}
Christopher~T. Hill and Elizabeth~H. Simmons.
\newblock {Strong Dynamics and Electroweak Symmetry Breaking}.
\newblock {\em Phys. Rept.}, 381:235--402, 2003.
\newblock [Erratum: Phys.Rept. 390, 553--554 (2004)].

\bibitem{Buttazzo:2013uya}
Dario Buttazzo, Giuseppe Degrassi, Pier~Paolo Giardino, Gian~F. Giudice,
  Filippo Sala, Alberto Salvio, and Alessandro Strumia.
\newblock {Investigating the near-criticality of the Higgs boson}.
\newblock {\em JHEP}, 12:089, 2013.

\bibitem{Bezrukov:2014ina}
Fedor Bezrukov and Mikhail Shaposhnikov.
\newblock {Why should we care about the top quark Yukawa coupling?}
\newblock {\em J. Exp. Theor. Phys.}, 120:335--343, 2015.

\bibitem{Carena:1993bs}
Marcela Carena, M.~Olechowski, S.~Pokorski, and C.~E.~M. Wagner.
\newblock {Radiative electroweak symmetry breaking and the infrared fixed point
  of the top quark mass}.
\newblock {\em Nucl. Phys. B}, 419:213--239, 1994.

\bibitem{Panico:2011pw}
Giuliano Panico and Andrea Wulzer.
\newblock {The Discrete Composite Higgs Model}.
\newblock {\em JHEP}, 09:135, 2011.

\bibitem{Matsedonskyi:2012ym}
Oleksii Matsedonskyi, Giuliano Panico, and Andrea Wulzer.
\newblock {Light Top Partners for a Light Composite Higgs}.
\newblock {\em JHEP}, 01:164, 2013.

\bibitem{Pomarol:2012qf}
Alex Pomarol and Francesco Riva.
\newblock {The Composite Higgs and Light Resonance Connection}.
\newblock {\em JHEP}, 08:135, 2012.

\bibitem{Bellazzini:2014yua}
Brando Bellazzini, Csaba Cs\'aki, and Javi Serra.
\newblock {Composite Higgses}.
\newblock {\em Eur. Phys. J. C}, 74(5):2766, 2014.

\bibitem{Panico:2015jxa}
Giuliano Panico and Andrea Wulzer.
\newblock {\em {The Composite Nambu-Goldstone Higgs}}, volume 913.
\newblock Springer, 2016.

\bibitem{Aad:2019mbh}
Georges Aad et~al.
\newblock {Combined measurements of Higgs boson production and decay using up
  to $80$ fb$^{-1}$ of proton-proton collision data at $\sqrt{s}=$ 13 TeV
  collected with the ATLAS experiment}.
\newblock {\em Phys. Rev. D}, 101(1):012002, 2020.

\bibitem{Aaboud:2018urx}
M.~Aaboud et~al.
\newblock {Observation of Higgs boson production in association with a top
  quark pair at the LHC with the ATLAS detector}.
\newblock {\em Phys. Lett. B}, 784:173--191, 2018.

\bibitem{Sirunyan:2018hoz}
Albert~M Sirunyan et~al.
\newblock {Observation of $\mathrm{t\overline{t}}$H production}.
\newblock {\em Phys. Rev. Lett.}, 120(23):231801, 2018.

\bibitem{Cepeda:2019klc}
M.~Cepeda et~al.
\newblock {Report from Working Group 2}: {Higgs Physics at the HL-LHC and
  HE-LHC}.
\newblock {\em CERN Yellow Rep. Monogr.}, 7:221--584, 2019.

\bibitem{Appelquist:1974tg}
Thomas Appelquist and J.~Carazzone.
\newblock {Infrared Singularities and Massive Fields}.
\newblock {\em Phys. Rev. D}, 11:2856, 1975.

\bibitem{Buchmuller:1985jz}
W.~Buchmuller and D.~Wyler.
\newblock {Effective Lagrangian Analysis of New Interactions and Flavor
  Conservation}.
\newblock {\em Nucl. Phys. B}, 268:621--653, 1986.

\bibitem{Grzadkowski:2010es}
B.~Grzadkowski, M.~Iskrzynski, M.~Misiak, and J.~Rosiek.
\newblock {Dimension-Six Terms in the Standard Model Lagrangian}.
\newblock {\em JHEP}, 10:085, 2010.

\bibitem{Azatov:2014jga}
Aleksandr Azatov, Christophe Grojean, Ayan Paul, and Ennio Salvioni.
\newblock {Taming the off-shell Higgs boson}.
\newblock {\em Zh. Eksp. Teor. Fiz.}, 147:410--425, 2015.

\bibitem{Englert:2014aca}
Christoph Englert and Michael Spannowsky.
\newblock {Limitations and Opportunities of Off-Shell Coupling Measurements}.
\newblock {\em Phys. Rev. D}, 90:053003, 2014.

\bibitem{Buschmann:2014sia}
Malte Buschmann, Dorival Goncalves, Silvan Kuttimalai, Marek Schonherr, Frank
  Krauss, and Tilman Plehn.
\newblock {Mass Effects in the Higgs-Gluon Coupling: Boosted vs Off-Shell
  Production}.
\newblock {\em JHEP}, 02:038, 2015.

\bibitem{Corbett:2015ksa}
Tyler Corbett, Oscar J.~P. Eboli, Dorival Goncalves, J.~Gonzalez-Fraile, Tilman
  Plehn, and Michael Rauch.
\newblock {The Higgs Legacy of the LHC Run I}.
\newblock {\em JHEP}, 08:156, 2015.

\bibitem{Goncalves:2017iub}
Dorival Goncalves, Tao Han, and Satyanarayan Mukhopadhyay.
\newblock {Off-Shell Higgs Probe of Naturalness}.
\newblock {\em Phys. Rev. Lett.}, 120(11):111801, 2018.
\newblock [Erratum: Phys.Rev.Lett. 121, 079902 (2018)].

\bibitem{Goncalves:2018pkt}
Dorival Gon\c{c}alves, Tao Han, and Satyanarayan Mukhopadhyay.
\newblock {Higgs Couplings at High Scales}.
\newblock {\em Phys. Rev. D}, 98(1):015023, 2018.

\bibitem{Goncalves:2020vyn}
Dorival Gon\c{c}alves, Tao Han, Sze Ching Iris~Leung, and Han Qin.
\newblock {Off-shell Higgs couplings in $H^*\to ZZ\to \ell\ell\nu\nu$}.
\newblock {\em Phys. Lett. B}, 817:136329, 2021.

\bibitem{Ellis:2020unq}
John Ellis, Maeve Madigan, Ken Mimasu, Veronica Sanz, and Tevong You.
\newblock {Top, Higgs, Diboson and Electroweak Fit to the Standard Model
  Effective Field Theory}.
\newblock 12 2020.

\bibitem{Ethier:2021bye}
Jacob~J. Ethier, Fabio Maltoni, Luca Mantani, Emanuele~R. Nocera, Juan Rojo,
  Emma Slade, Eleni Vryonidou, and Cen Zhang.
\newblock {Combined SMEFT interpretation of Higgs, diboson, and top quark data
  from the LHC}.
\newblock 4 2021.

\bibitem{Brivio:2019ius}
Ilaria Brivio, Sebastian Bruggisser, Fabio Maltoni, Rhea Moutafis, Tilman
  Plehn, Eleni Vryonidou, Susanne Westhoff, and C.~Zhang.
\newblock {O new physics, where art thou? A global search in the top sector}.
\newblock {\em JHEP}, 02:131, 2020.

\bibitem{Biekotter:2018jzu}
Anke Biek\"otter, Dorival Gon\c{c}alves, Tilman Plehn, Michihisa Takeuchi, and
  Dirk Zerwas.
\newblock {The global Higgs picture at 27 TeV}.
\newblock {\em SciPost Phys.}, 6(2):024, 2019.

\bibitem{Maltoni:2016yxb}
Fabio Maltoni, Eleni Vryonidou, and Cen Zhang.
\newblock {Higgs production in association with a top-antitop pair in the
  Standard Model Effective Field Theory at NLO in QCD}.
\newblock {\em JHEP}, 10:123, 2016.

\bibitem{ATLAS-CONF-2020-027}
{A combination of measurements of Higgs boson production and decay using up to
  $139$ fb$^{-1}$ of proton--proton collision data at $\sqrt{s}=$ 13 TeV
  collected with the ATLAS experiment}.
\newblock Technical report, CERN, Geneva, Aug 2020.

\bibitem{CMS:2018jcg}
Albert~M Sirunyan et~al.
\newblock {Measurement of the top quark polarization and $\mathrm{t\bar{t}}$
  spin correlations using dilepton final states in proton-proton collisions at
  $\sqrt{s} =$ 13 TeV}.
\newblock {\em Phys. Rev. D}, 100(7):072002, 2019.

\bibitem{Liu:2017dsz}
Da~Liu, Ian Low, and Carlos E.~M. Wagner.
\newblock {Modification of Higgs Couplings in Minimal Composite Models}.
\newblock {\em Phys. Rev. D}, 96(3):035013, 2017.

\bibitem{Banerjee:2021qhr}
Avik Banerjee, Sayan Dasgupta, and Tirtha~Sankar Ray.
\newblock {Chasing the Higgs shape at HL-LHC}.
\newblock 5 2021.

\bibitem{Punjabi:2015bba}
V.~Punjabi, C.~F. Perdrisat, M.~K. Jones, E.~J. Brash, and C.~E. Carlson.
\newblock {The Structure of the Nucleon: Elastic Electromagnetic Form Factors}.
\newblock {\em Eur. Phys. J. A}, 51:79, 2015.

\bibitem{Alwall:2014hca}
J.~Alwall, R.~Frederix, S.~Frixione, V.~Hirschi, F.~Maltoni, O.~Mattelaer,
  H.~S. Shao, T.~Stelzer, P.~Torrielli, and M.~Zaro.
\newblock {The automated computation of tree-level and next-to-leading order
  differential cross sections, and their matching to parton shower
  simulations}.
\newblock {\em JHEP}, 07:079, 2014.

\bibitem{Degrande:2020evl}
C\'eline Degrande, Gauthier Durieux, Fabio Maltoni, Ken Mimasu, Eleni
  Vryonidou, and Cen Zhang.
\newblock {Automated one-loop computations in the SMEFT}.
\newblock 8 2020.

\bibitem{Goldouzian:2020ekx}
Reza Goldouzian, Jeong~Han Kim, Kevin Lannon, Adam Martin, Kelci Mohrman, and
  Andrew Wightman.
\newblock {Matching in $pp \to t \bar{t} W/Z/h +$ jet SMEFT studies}.
\newblock 12 2020.

\bibitem{Sjostrand:2014zea}
Torbj\"orn Sj\"ostrand, Stefan Ask, Jesper~R. Christiansen, Richard Corke,
  Nishita Desai, Philip Ilten, Stephen Mrenna, Stefan Prestel, Christine~O.
  Rasmussen, and Peter~Z. Skands.
\newblock {An introduction to PYTHIA 8.2}.
\newblock {\em Comput. Phys. Commun.}, 191:159--177, 2015.

\bibitem{Skands:2014pea}
Peter Skands, Stefano Carrazza, and Juan Rojo.
\newblock {Tuning PYTHIA 8.1: the Monash 2013 Tune}.
\newblock {\em Eur. Phys. J. C}, 74(8):3024, 2014.

\bibitem{Artoisenet:2012st}
Pierre Artoisenet, Rikkert Frederix, Olivier Mattelaer, and Robbert Rietkerk.
\newblock {Automatic spin-entangled decays of heavy resonances in Monte Carlo
  simulations}.
\newblock {\em JHEP}, 03:015, 2013.

\bibitem{Harland-Lang:2014zoa}
L.~A. Harland-Lang, A.~D. Martin, P.~Motylinski, and R.~S. Thorne.
\newblock {Parton distributions in the LHC era: MMHT 2014 PDFs}.
\newblock {\em Eur. Phys. J. C}, 75(5):204, 2015.

\bibitem{Buckley:2015vsa}
Matthew~R. Buckley and Dorival Goncalves.
\newblock {Boosting the Direct CP Measurement of the Higgs-Top Coupling}.
\newblock {\em Phys. Rev. Lett.}, 116(9):091801, 2016.

\bibitem{Cacciari:2011ma}
Matteo Cacciari, Gavin~P. Salam, and Gregory Soyez.
\newblock {FastJet User Manual}.
\newblock {\em Eur. Phys. J. C}, 72:1896, 2012.

\bibitem{Butterworth:2008iy}
Jonathan~M. Butterworth, Adam~R. Davison, Mathieu Rubin, and Gavin~P. Salam.
\newblock {Jet substructure as a new Higgs search channel at the LHC}.
\newblock {\em Phys. Rev. Lett.}, 100:242001, 2008.

\bibitem{Plehn:2009rk}
Tilman Plehn, Gavin~P. Salam, and Michael Spannowsky.
\newblock {Fat Jets for a Light Higgs}.
\newblock {\em Phys. Rev. Lett.}, 104:111801, 2010.

\bibitem{CERN-LHCC-2017-021}
{Technical Design Report for the ATLAS Inner Tracker Pixel Detector}.
\newblock Technical Report CERN-LHCC-2017-021. ATLAS-TDR-030, CERN, Geneva, Sep
  2017.

\bibitem{CMS-PAS-TOP-19-001}
{Using associated top quark production to probe for new physics within the
  framework of effective field theory}.
\newblock Technical report, CERN, Geneva, 2020.

\bibitem{Brehmer:2019xox}
Johann Brehmer, Felix Kling, Irina Espejo, and Kyle Cranmer.
\newblock {MadMiner: Machine learning-based inference for particle physics}.
\newblock {\em Comput. Softw. Big Sci.}, 4(1):3, 2020.

\bibitem{Aaboud:2017rss}
Morad Aaboud et~al.
\newblock {Search for the standard model Higgs boson produced in association
  with top quarks and decaying into a $b\bar{b}$ pair in $pp$ collisions at
  $\sqrt{s}$ = 13 TeV with the ATLAS detector}.
\newblock {\em Phys. Rev. D}, 97(7):072016, 2018.

\bibitem{Jezo:2018yaf}
Tom\'a\v{s} Je\v{z}o, Jonas~M. Lindert, Niccolo Moretti, and Stefano Pozzorini.
\newblock {New NLOPS predictions for $t \bar{t} +b$ -jet production at the
  LHC}.
\newblock {\em Eur. Phys. J. C}, 78(6):502, 2018.

\bibitem{Denner:2020orv}
Ansgar Denner, Jean-Nicolas Lang, and Mathieu Pellen.
\newblock {Full NLO QCD corrections to off-shell
  tt\textasciimacron{}bb\textasciimacron{} production}.
\newblock {\em Phys. Rev. D}, 104(5):056018, 2021.

\bibitem{Bevilacqua:2021cit}
Giuseppe Bevilacqua, Huan-Yu Bi, Heribertus~Bayu Hartanto, Manfred Kraus,
  Michele Lupattelli, and Malgorzata Worek.
\newblock {$ t\overline{t}b\overline{b} $ at the LHC: on the size of
  corrections and b-jet definitions}.
\newblock {\em JHEP}, 08:008, 2021.

\end{thebibliography}

\end{document}